\documentclass[journal=jacsat,manuscript=article]{achemso}

\usepackage[version=3]{mhchem} % Formula subscripts using \ce{}
\usepackage{amsmath}
\usepackage{amssymb}
\usepackage{hyperref}
\usepackage{nameref}
\usepackage{float}
\usepackage{fontawesome5} % 提供 GitHub 和 Hugging Face 图标
\usepackage{hyperref}      % 提供超链接功能

\usepackage{fancyhdr}
\usepackage{graphicx}
\usepackage{placeins}
\usepackage{chngcntr}  % 用于修改计数器
\usepackage{booktabs}
\usepackage{natbib}
\author{Xin-Yu Lu}
\affiliation[]
{College of Chemistry and Chemical Engineering, Xiamen University, Xiamen, China}
\alsoaffiliation[]
{Shanghai Innovation Institute, Shanghai, China}
\alsoaffiliation[]
{Tan Kah Kee Innovation Laboratory, Xiamen, China}

\author{De-Yi Lin}
\affiliation{Institute of Artificial Intelligence, Xiamen University, Xiamen 361005, China}

\author{Tong Zhu}
\affiliation[]{Shanghai Innovation Institute, Shanghai, China}
\alsoaffiliation[East China Normal University]
{Department of Chemistry, East China Normal University, Shanghai, China}

\author{Bin Ren}
\affiliation[]
{College of Chemistry and Chemical Engineering, Xiamen University, Xiamen, China}
\alsoaffiliation[]
{Tan Kah Kee Innovation Laboratory, Xiamen, China}
\alsoaffiliation{Institute of Artificial Intelligence, Xiamen University, Xiamen 361005, China}

\author{Hao Ma}
\affiliation[]
{College of Chemistry and Chemical Engineering, Xiamen University, Xiamen, China}
\alsoaffiliation[]
{Tan Kah Kee Innovation Laboratory, Xiamen, China}
\alsoaffiliation{Institute of Artificial Intelligence, Xiamen University, Xiamen 361005, China}
\email{oaham@xmu.edu.cn}

\author{Guo-Kun Liu}
\affiliation[]
{College of the Environment and Ecology, Xiamen University, Xiamen, China}
\email{guokunliu@xmu.edu.cn}

\title[]
  {Vib2Conf: AI-driven discrimination of molecular conformations from vibrational spectra}

\begin{document}
\thispagestyle{fancy}
\begin{tocentry}
\begin{center}
  \includegraphics[width=1.00\textwidth]{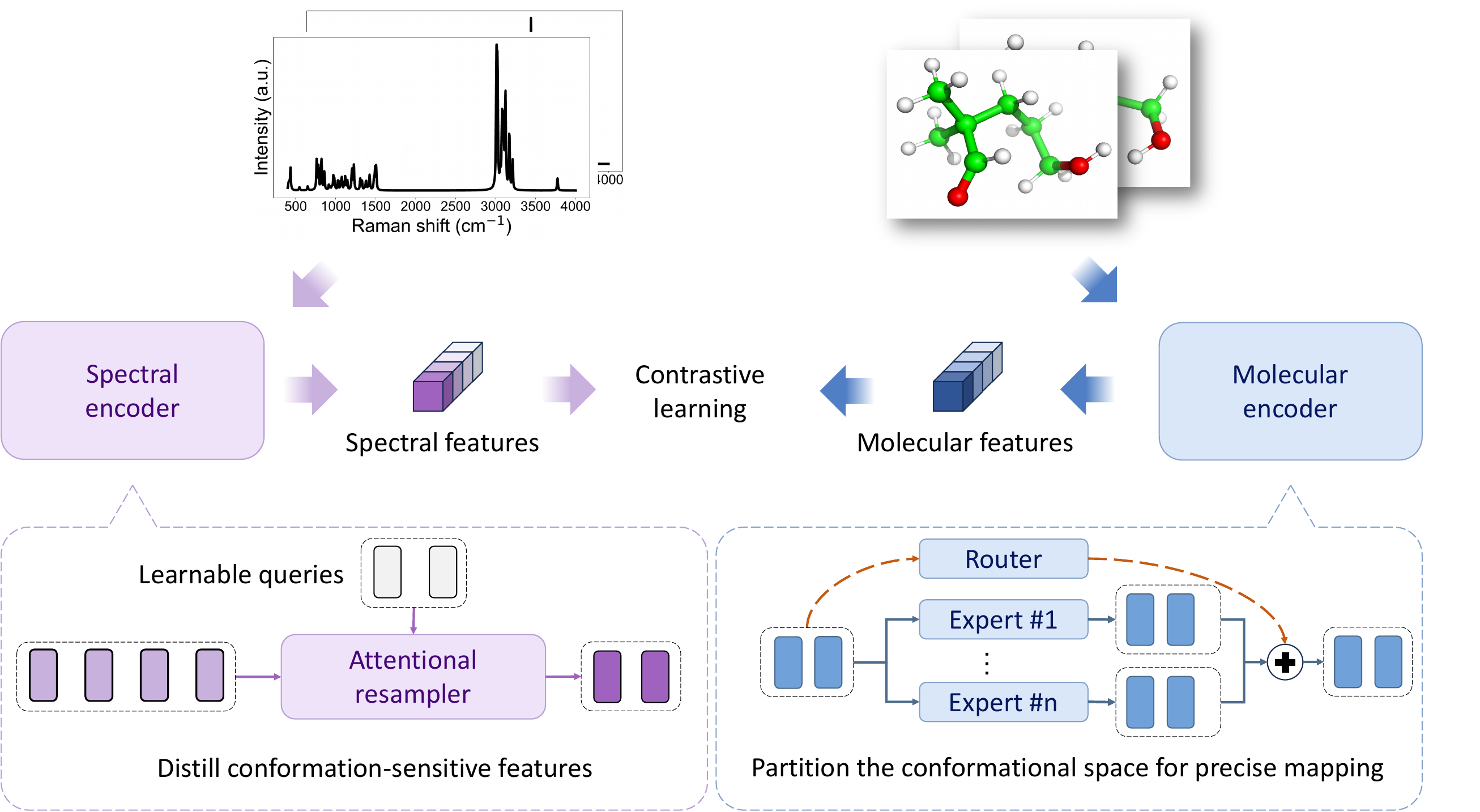}
\end{center}
\end{tocentry}

\begin{abstract}
  Retrieving or generating two-dimensional molecular structures on the basis of vibrational spectra has been well demonstrated via deep learning models. However, deciphering three-dimensional molecular conformations is still challenging, primarily due to spectral ambiguities caused by conformational heterogeneity, which are difficult to resolve. To address this limitation, we propose Vib2Conf, a deep learning model directly discriminating 3D molecular conformations from vibrational spectra. We implement an attentional resampler to distill conformation-sensitive features from sparse spectral signals, and integrate Mixture-of-Experts (MoE) to partition the conformational space for precise geometric mapping. These modules enable Vib2Conf to achieve state-of-the-art top-1 recall exceeding 95\% on traditional spectrum-structure benchmarks, including QM9S, VB-Mols, and QMe14S. More importantly, Vib2Conf can discriminate near-isomeric conformers with a top-1 recall of 82.06\% on VB-Confs test set, where conformational isomers differ by a root-mean-square deviation (RMSD) of only ~1 Å. In general, Vib2Conf is a promising method for fine-grained spectrum-to-conformation analysis.
\end{abstract}

\begin{center}
    \vspace{0.5em} 
    \hypersetup{colorlinks=true, urlcolor=black} 

    \begin{tabular}{l}
        \href{https://github.com/X1nyuLu/vib2conf}{%
            \faGithub\ \small https://github.com/X1nyuLu/vib2conf%
        } 
    \end{tabular}
    \vspace{0.5em}
\end{center}

\pagestyle{empty}
\section{Introduction}
Molecular conformation governs the function and reactivity of chemical and biological systems,\cite{ref1, ref2, ref3, ref4} yet its precise determination remains a fundamental challenge due to the inherent dynamic and heterogeneous nature of molecular states. Conventional bulk measurements generally yield ensemble-averaged observables, masking the underlying conformational heterogeneity within molecular populations.\cite{ref5} Consequently, most characterization strategies are still limited to simulated data, lacking the robustness required for direct interpretation of real-world experimental signatures.\cite{ref6, ref7} In contrast, single-molecule experiments have unveiled the transformative potential of vibrational spectroscopy as a real-time probe, where vibrational signatures serve as an exquisite readout for even the most minute structural fluctuations.\cite{ref8, ref9, ref10} However, extending this proof-of-principle capability to broadly applicable conformational analysis requires not only versatile and ultra-precise characterization methods for the large-scale acquisition of high-fidelity experimental data, but also a robust spectrum-to-structure inference method to bridge the informational gap between 1D spectral signals and 3D conformational geometries.

Realization of such spectrum-to-structure task is hindered by two major challenges: First, current models (including Vib2Mol\cite{ref11}, VibraCLIP\cite{ref12} and SMEN\cite{ref13}) normally overlook a fundamental fact: there is a large gap in information density (entropy) between spectra and molecules. Specifically, a 3D molecular conformation provides a dense, complete structural description, whereas a vibrational spectrum consists of pixel-wise intensities. Within this 1D representation, spectral information is scattered across hundreds of individual data points, many of which are redundant or obscured by peak overlap. This disparity in information density necessitates architectural innovations to facilitate effective distillation of spectral features and the comprehensive extraction of conformational features. Second, a key bottleneck lies in the scarcity of benchmarks that resolve conformational degeneracy. Currently, there are few high-quality databases that map vibrational spectra to individual conformations, regardless of whether they come from experiments or quantum simulations. Without such benchmarks, it is impossible to evaluate whether a deep learning model can truly deconvolve conformational degeneracy, where closely related conformers with minute deviations yield nearly indistinguishable vibrational signatures.

Therefore, we proposed Vib2Conf, a deep learning model capable of accurately retrieving molecular structures and their corresponding 3D conformations directly from vibrational spectra. Vib2Conf addresses spectral redundancy through an attentional resampler which distills conformation-sensitive features from sparse signals. Simultaneously, to capture the intricate geometric nuances of 3D space, we integrate Mixture-of-Experts (MoE) into an equivariant backbone, enabling a "divide-and-conquer" strategy for precise conformational mapping. As a result, Vib2Conf achieves state-of-the-art performance on traditional spectrum-structure retrieval benchmarks, including QM9S\cite{ref14}, VB-mols\cite{ref11} and QMe14S\cite{ref15}. Furthermore, to evaluate model performance in deciphering subtle structural variations with a RMSD of ~ 1 Å, we introduced VB-Confs, an extension of our previous vibrational spectrum-to-structure benchmark series. Vib2Conf demonstrates a superior top-1 recall of 82.06\% in discriminating between near-isomeric conformers. This highlights its ability to resolve structural differences at an unprecedented resolution.

\section{Methods}

\subsection{Framework of Vib2Conf}
\begin{figure}[htbp]
  \centering
  \includegraphics[width=1\textwidth]{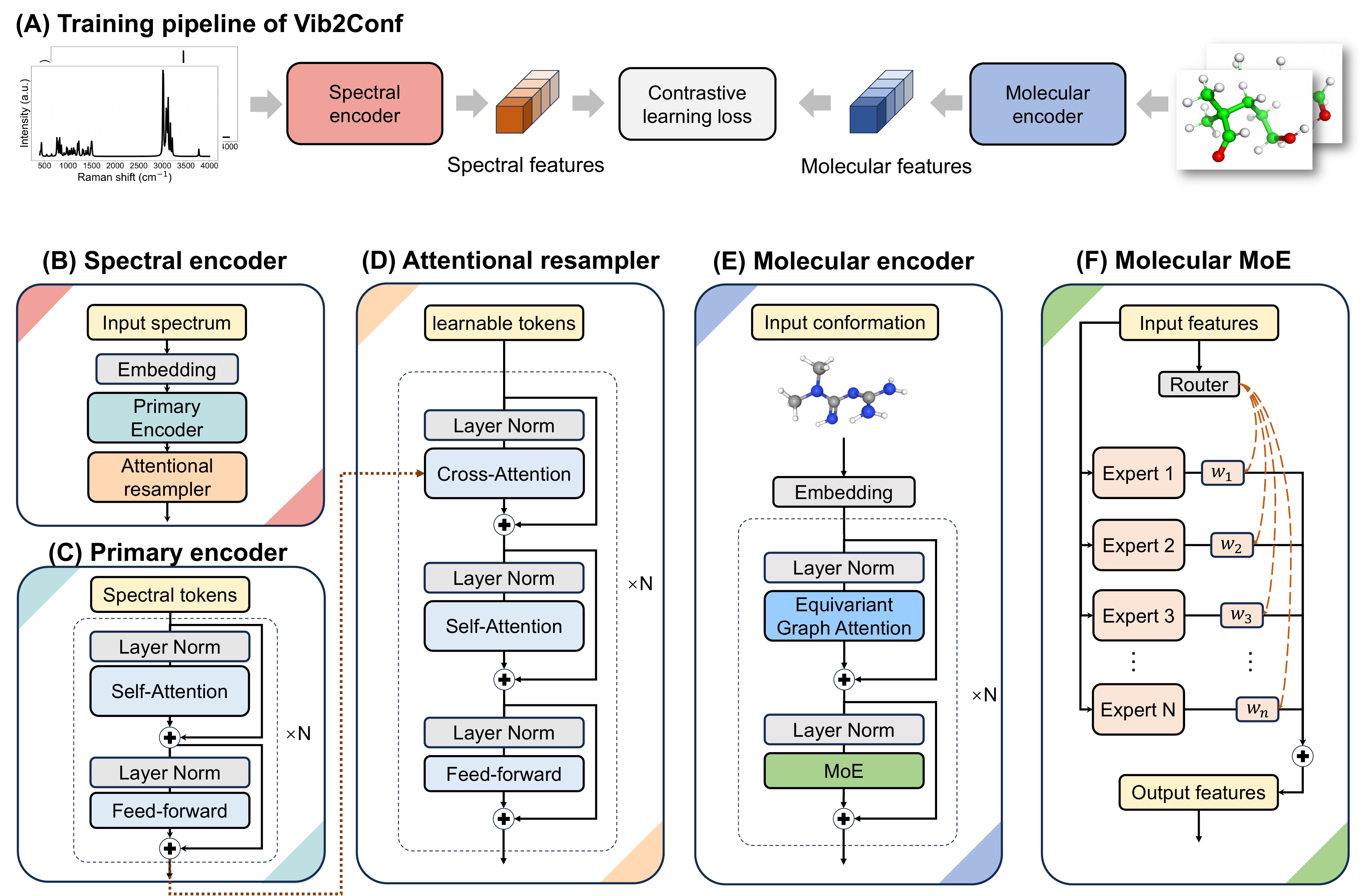}
  \caption{Schematic illustration of the Vib2Conf architecture. (A) The training pipeline of Vib2Conf: the core architecture consists of a spectral encoder and a molecular encoder, which project vibrational spectra and 3D molecular conformations into a unified latent space via contrastive learning for bidirectional retrieval. (B) Spectral encoder: the workflow involves encoding raw spectra into spectral tokens, followed by primary encoder and further resampling. (C) Primary encoder: a standard Transformer block utilized for preliminary spectral feature extraction. (D) Attentional resampler: a distillation module compressing 128 initial tokens into 64 refined learnable tokens through cross-attention to eliminate informational redundancy. (E) Molecular encoder: an Equiformer-based backbone optimized for extracting 3D conformational features. (F) Mixture of Experts (MoE): the specialized feed-forward network within the molecular encoder, where a router dynamically assigns weights to multiple experts to capture diverse structural nuances.}
  \label{fig:1}
\end{figure}

As illustrated in Figure \ref{fig:1}A, we adopt a dual-tower framework consisting of a spectral encoder for processing vibrational spectra (IR and Raman) and a molecular encoder for extracting conformational features. Through contrastive learning, spectral and molecular features are projected and aligned within a unified latent space, enabling spectrum-conformation retrieval.

The spectral encoder architecture is detailed in Figure \ref{fig:1}B. Initially, each spectrum is partitioned into 128 contiguous patches and projected into a sequence of tokens via an embedding layer. Subsequently, these tokens undergo feature extraction via a primary encoder and an attentional resampler. The primary encoder (Figure \ref{fig:1}C), a standard Transformer encoder comprising self-attention and feed-forward layers, is employed to extract initial spectral representations. Following this, the attentional resampler performs feature distillation (Figure \ref{fig:1}D), in which a set of 64 learnable tokens is initialized. The features from the primary encoder are refined into these learnable tokens via cross-attention (see details in supporting information), followed by self-attention and feed-forward blocks to generate the final refined spectral tokens.

The architecture of the molecular encoder (Figure \ref{fig:1}E) is modified on the basis of Equiformer\cite{ref16}. Specifically, the feed-forward neural networks are substituted by mixture of experts (MoE), which incorporates multiple individual linear layers (experts) to capture diverse molecular features (Figure \ref{fig:1}F). The final output is a weighted linear combination of these experts, with the contribution of each expert dynamically determined by a router mechanism.

\FloatBarrier

\subsection{Training strategy and loss functions}
During the training phase, the overall loss function, as formulated in Equation \ref{eqn:loss_total}, is defined as the weighted sum of the symmetric contrastive loss\cite{ref17} (Equation \ref{eqn:s1}, \ref{eqn:s2}) and the load-balancing loss. The former is employed to minimize the similarity errors between the spectrum and related conformation of the same molecule, while the latter ensures a balanced workload across the MoE module. To ensure that the load-balancing constraint does not adversely interfere with the primary convergence of the retrieval task\cite{ref18}, the balancing hyperparameter is set to $\alpha=0.0001$ (see details in supporting information). This minimal weight allows the model to prioritize feature alignment while maintaining a sufficiently uniform distribution of expert utilization.

\begin{equation}
    L_{total}=L_{contrastive}+ \alpha \times L_{balance}
    \label{eqn:loss_total}
\end{equation}

The load-balancing loss is defined as the square of the coefficient of variation ($CV^2$) of the expert weights. Let $W \in \mathbb{R}^{B\times{N}}$ denote the routing weights for a mini-batch of size $B$ across $N$ experts. The importance of each expert $j$ is accumulated over the batch as $w_j=\sum_{i=1}^{B}W_{i,j}$. The balancing loss is formulated as Equation \ref{eqn:loss_balance}. This loss term penalizes large variances in expert utilization, and incentivizes the model to explore and utilize the full capacity of the expert pool.

\begin{equation}
    L_{balance}=N\cdot\frac{\sum_{j=1}^{N}w_j^2}{\left(\sum_{j=1}^{N}w_j\right)^2}-1=N\cdot\frac{\sum_{j=1}^{N}\left(\sum_{i=1}^{B}W_{i,j}\right)^2}{\left(\sum_{j=1}^{N}\sum_{i=1}^{B}W_{i,j}\right)^2}-1
    \label{eqn:loss_balance}
\end{equation}

During the testing or inference phase, it requires only the computation of the dot product between the feature vector of the query spectrum and the pre-computed feature vectors of the molecular library. The resulting similarity scores are then ranked, and the top-k candidates are retrieved as the final results. Detailed hyper-parameters about model and training are recorded in Table \ref{tab:hyperparams}.

\subsection{Datasets and benchmarks}

Two distinct categories of benchmarks are used to evaluate model performance. The first one focuses on the fundamental spectrum-to-structure retrieval task, which involves identifying the matching molecular identity from a database on the basis of a query spectrum. To this end, we utilize three open-source datasets, QM9S, VB-Mols, and QMe14S, which provide a diverse collection of unique molecule-Raman-IR triplets derived from quantum chemical simulations. These datasets represent varying degrees of chemical complexity. QM9S contain only C, H, O, N, and F, representing the most constrained chemical space. VB-Mols exhibits slightly higher complexity (including C, H, O, N, F, Cl, Br, S, P, and Si), while QMe14S is the most expansive, encompassing C, H, O, N, F, Cl, Br, S, P, Si, B, Al, As and Se.

The second category focuses on the more challenging task of spectrum-to-conformation retrieval. In this scenario, the model must not only identify the molecule but also resolve its specific 3D spatial arrangement by discriminating between multiple stable conformers of the same chemical identity. This is conducted using our in-house developed benchmark, ViBench-Confs (VB-Confs), an extension of our previous vibrational spectrum-to-structure benchmark series. Specifically designed to test structural resolution, this dataset contains 20,703 molecules derived from quantum mechanical simulations, with each molecule featuring 10 distinct stable conformations and their corresponding simulated Raman/IR spectra. The distribution of the RMSD for conformational isomers is illustrated in Figure S4. The complete dataset encompasses a broad conformational diversity, ranging from nearly identical conformers (RMSD $<$ 0.1 Å) to significantly distinct conformational isomers (RMSD $>$ 2 Å), with a mean distribution of 1.36 Å. Given that 1 Å is generally regarded as the conventional threshold for distinguishing between individual conformational states, VB-Confs possesses the requisite resolution to evaluate the capacity of a model to resolve both subtle structural fluctuations and major geometric transitions.\cite{ref19}

The construction of VB-Confs was based on the GEOM-QM9\cite{ref20} repository. 20,703 molecules were randomly selected, along with the 10 lowest-energy conformations for each. Using these conformations as initial coordinates, we performed geometry optimization and vibrational spectra calculations via Density Functional Theory (DFT). Unless otherwise specified, all quantum chemical calculations were conducted using the Gaussian 09 program\cite{ref21}. Geometries were optimized utilizing the B3LYP-D3BJ functional paired with a 6-311+G** basis set. Harmonic frequency calculations were subsequently performed at the same level of theory on the optimized geometries. To standardize the data for subsequent modeling, all spectra were cropped to the range of 400-4000 $cm^{-1}$, and unified to a fixed resolution of 3.51 $cm^{-1}$, resulting in a consistent input size of 1024 pixels.

During the partitioning of the training, validation, and test sets, we employed distinct strategies for spectrum-structure retrieval and spectrum-conformation retrieval. For the former, a random sampling approach was adopted. In contrast, for spectrum-conformation retrieval, we implemented splits on the basis of molecular identity. This rigorous partitioning ensures that all conformers associated with a specific chemical structure are assigned exclusively to either the training, validation, or test set, preventing any data leakage between sets. Consequently, the trained model must generalize to entirely novel molecules and retrieve the most relevant conformation from the library on the basis of a query spectrum. For all datasets, the proportions for the training, validation, and test sets were maintained at a ratio of 85\%:5\%:10\%.

\subsection{Baseline models}
We evaluated the spectrum-structure retrieval performance of Vib2Conf against three baseline models (Vib2Mol, VibraCLIP and SMEN) on the QM9S, VB-Mols, and QMe14S datasets. On the basis of molecular representation, these models are categorized into three classes: the 1D SMILES-based Vib2Mol, the 2D graph-based VibraCLIP, and the 3D conformation-based SMEN and Vib2Conf. These model all leverage contrastive learning and achieve robust performance in spectrum-structure retrieval.

\section{Results}
\subsection{Benchmarking Vib2Conf on traditional spectrum-structure retrieval}
\begin{figure}[htbp]
\centering
\includegraphics[width=1\textwidth]{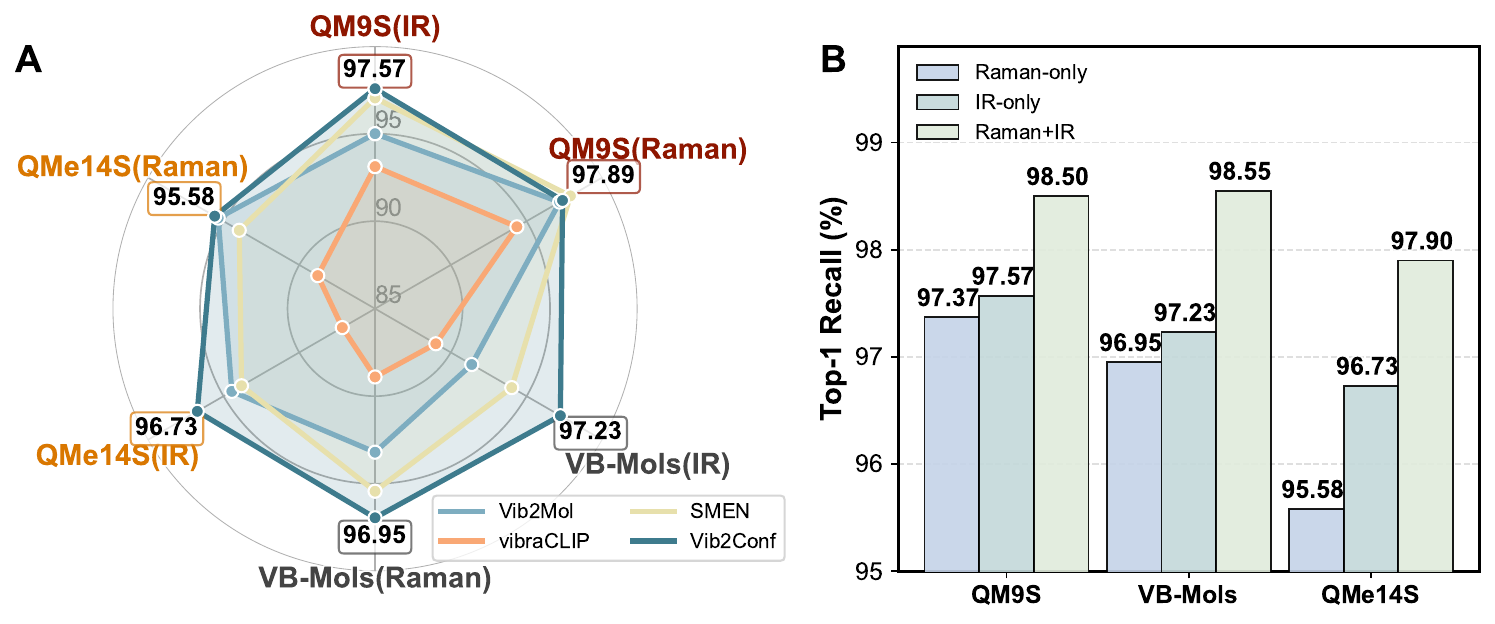}
\caption{(A) Comparative evaluation of spectrum-structure retrieval. Performance of Vib2Conf is compared against baseline models (Vib2Mol, VibraCLIPand SMEN) and achieves state-of-the-art performance in 5 out of 6 benchmarks. (B) Impact of multimodal fusion on retrieval performance. Detailed numerical data for these evaluations are provided in Table \ref{tab:qm9s}, \ref{tab:vbmols}, and \ref{tab:qme14s}.}
\label{fig:2}
\end{figure}

Figure \ref{fig:2}A illustrates the state-of-the-art performance of Vib2Conf against three baseline models (Vib2Mol, VibraCLIP and SMEN) across Raman and IR spectral inputs (detailed data in Table \ref{tab:qm9s}, \ref{tab:vbmols}, \ref{tab:qme14s}). On the relatively simple QM9S dataset, the performance gap among these models is marginal, with all Recall@1 values converging at approximately 97\%. However, the advantages of Vib2Conf become increasingly pronounced as dataset complexity grows. On the more challenging VB-Mols benchmark, Vib2Conf achieves a Recall@1 of 97.23\% for IR, outperforming SMEN \(94.02\%\), Vib2Mol \(91.38\%\), and VibraCLIP \(89.01\%\). This leadership is further solidified on the most complex QMe14S dataset, where Vib2Conf reaches 95.58\% for IR, surpassing Vib2Mol \(94.45\%\), SMEN \(93.81\%\), and VibraCLIP \(87.16\%\).

Further, the integration of Raman and IR spectral information yields a significant performance enhancement for Vib2Conf (Figure \ref{fig:2}B and Table \ref{tab:multi-modal}). For instance, an improvement is from 97.37\% (Raman) to 98.50\% (Raman+IR) on the relatively simple QM9S dataset, and from 95.58\% (Raman-only) to 97.90\% (Raman+IR) on the more challenging QMe14S dataset. This tendency suggests that additional spectral information input provides greater value as task complexity increases.

\FloatBarrier

\subsection{Benchmarking Vib2Conf on spectrum-conformation retrieval}
The spectrum-conformation retrieval necessitates overcoming two primary challenges: mitigating interference from structurally similar molecules and resolving the subtle differences between conformers of the same molecule.

\begin{figure}[htbp]
  \centering
  \includegraphics[width=1\textwidth]{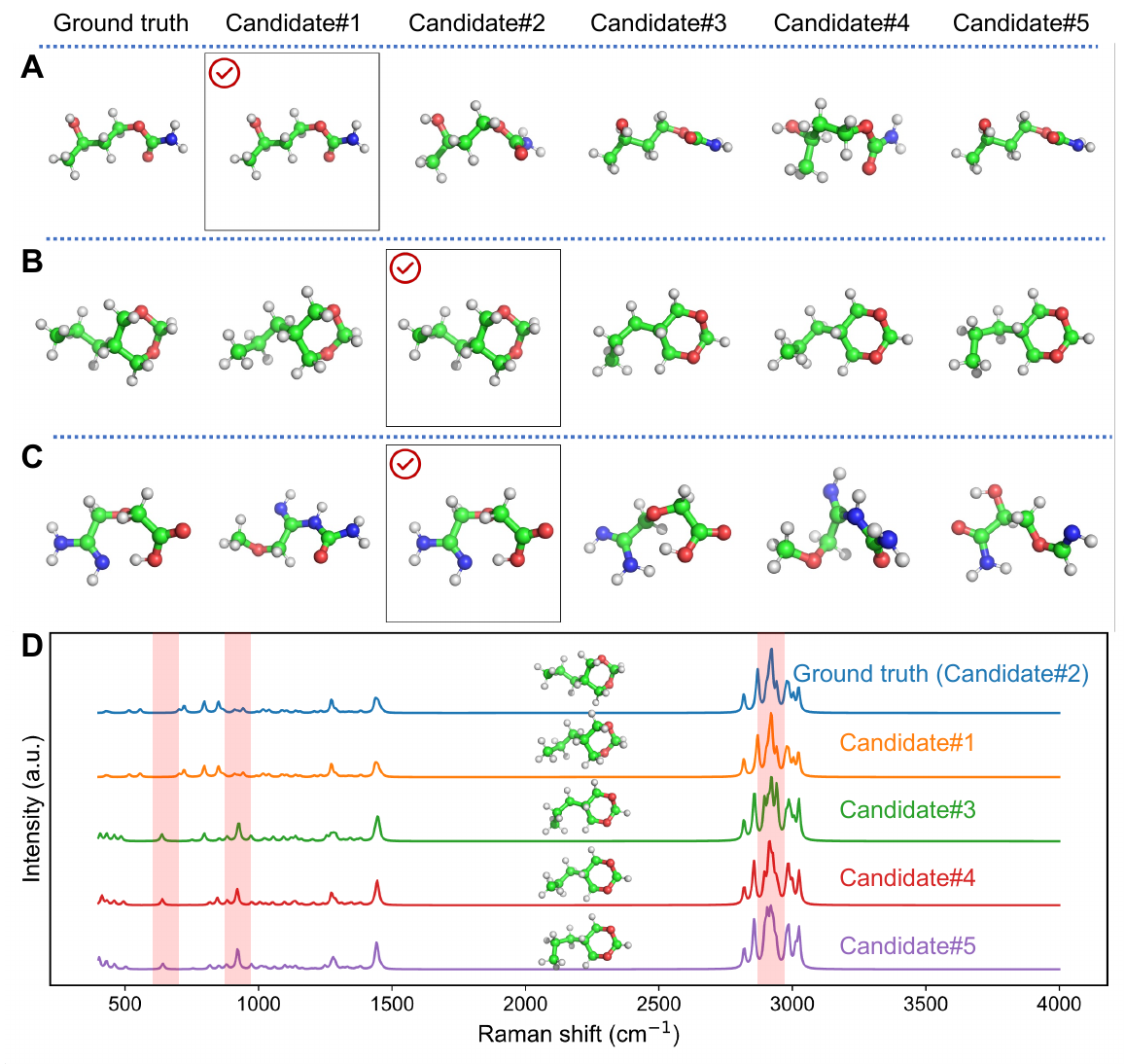}
  \caption{Schematic of retrieval outcomes in spectrum-conformation Retrieval on the VB-Confs Benchmark. (A) Perfectly correct, where the retrieved conformation matches the ground truth. (B) Partially correct, where the model identifies the correct molecular species (same SMILES) but selects an incorrect spatial conformer. (C) Incorrect, where the retrieved result corresponds to a different molecular species. (D) Representative case of intra-molecular conformational ambiguity. Comparison of Raman spectra for various conformers of a single molecule. While conformers \#3, \#4, and \#5 exhibit distinct spectral fingerprints at 640 and 925 $cm^{-1}$ relative to the ground truth, the "partially correct" candidate (Candidate \#1) shows high spectral similarity, with differences limited to minor relative intensity variations in the 2850-3000 $cm^{-1}$ C-H stretching region.}
  \label{fig:3}
\end{figure}

As illustrated in Figure \ref{fig:3}A-C, the results of spectrum-conformation retrieval on the VB-Confs benchmark are classified into three categories. Given Raman spectrum, the first category is “perfectly correct” constituting 81.72\% of the cases, where the retrieved and target 3D conformation are identical. The second category is “partially correct” \(18.22\%\), referring to instances where the model successfully identifies the correct molecular species but retrieves an alternative conformer rather than the target conformation. Finally, "totally incorrect" results represent a negligible 0.06\%, denoting cases where the retrieved molecule belongs to an entirely different species. Such distribution indicates that Vib2Conf effectively handles interference from similar molecules, with the majority of errors arising from intra-molecular conformational ambiguity.

Furthermore, taking the “partially correct” case of 2-propyl-1,3-dioxane in Figure 3B as an example, we visualize the corresponding Raman spectrum of ground truth (conformations \#2) and mis-retrieved candidates (conformations \#1, \#3, \#4, and \#5). As illustrated in Figure 3D, the spectrum of conformations \#1 is almost indistinguishable to that of the ground truth, and the difference spectrum between them shows only minor relative intensity differences around 2850-3000 $cm^{-1}$ C-H stretching region (Figure \ref{fig:s5}). In contrast, conformations \#3, \#4, and \#5 exhibit distinct spectral signatures not only around this region but also that around 640 and 925 $cm^{-1}$. This implies that the model's retrieval errors in “partially correct” may stem from the extreme spectral similarity among near-isomeric conformers.

\begin{figure}[htbp]
  \centering
  \includegraphics[width=1\textwidth]{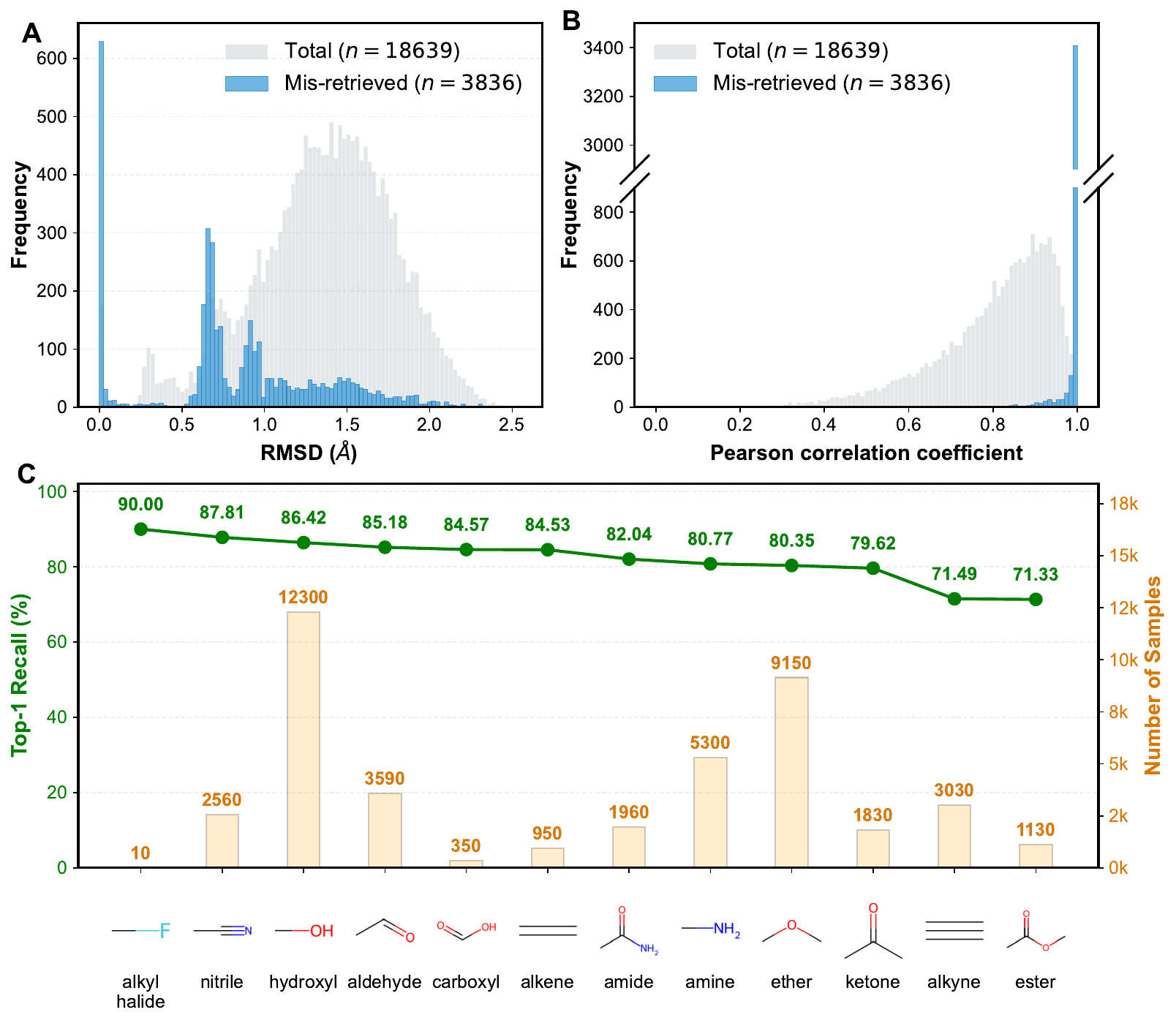}
  \caption{Statistical analysis of spectrum-conformation retrieval performance and error attribution. (A) Distribution of the root mean square deviation (RMSD) between retrieved conformations and ground truth for partially correct cases. (B) Distribution of the correlation coefficient $\rho$ between retrieved and target spectra. (C) Recall@1 performance across diverse functional groups.}
  \label{fig:4}
\end{figure}

We quantitatively visualized the distribution of conformational deviations by the root mean square deviation (RMSD) between the retrieved conformations and the ground truth (Figure \ref{fig:4}A). It is observed that the mis-retrieved candidates exhibit marginal geometric deviations from the ground truth. Specifically, the mean RMSD for these cases is only 0.84 Å, which falls well below the 1 Å threshold typically used to define identical or near-isomeric conformations\cite{ref22}. Furthermore, we analyzed the distribution of spectral similarity using the Pearson correlation coefficient $\rho$ on test set. As illustrated in Figure \ref{fig:4}B, the gray histogram represents the baseline similarity distribution among all 10 candidate conformations for each molecule. Meanwhile, the blue histogram highlights the similarity between the mis-retrieved candidates and the ground-truth targets. Notably, the blue distribution is heavily skewed toward 0.99, indicating that the failure cases predominantly occur when the spectral profiles of the candidates are hardly distinguishable from the ground truth.

To further understand the causes of conformational errors, we conducted a statistical analysis of performance across various functional groups, as spectral discriminability is intrinsically linked to chemical environments. As illustrated in Figure \ref{fig:4}C, functional groups situated at the terminal positions of molecular chains consistently exhibit superior Recall@1 values. A representative comparison is observed between the nitrile \(87.81\%\) and alkyl \(71.49\%\) groups. While both groups possess highly distinctive characteristic peaks that are readily resolved in the vibrational spectrum, their structural roles differ significantly. The nitrile group typically functions as a terminal feature with restricted geometric variability. In contrast, alkyl chains frequently serve as internal bridges connecting diverse substituents at both ends. This structural positioning introduces a higher degree of freedom and a more expansive conformational distribution, which complicates the retrieval task and leads to the observed performance disparity.

Interestingly, as illustrated by Table \ref{tab:confs}, the Recall@1 achieved with Raman spectral inputs \(81.72\%\) significantly outperforms that achieved with IR spectral inputs \(74.52\%\) in spectrum–conformation retrieval. Moreover, multimodal fusion combining Raman and IR signals only marginally improves the performance compared to the Raman-only baseline, increasing the Recall@1 from 81.72\% to 82.06\%. The performance gap resulting from the choice of input signal may be ascribed to the underlying physical tensors. In specific, IR spectral intensities are determined by the first-order derivatives of the dipole moment (a rank-1 tensor), whereas Raman spectral intensities depend on the corresponding derivatives of the molecular polarizability (a rank-2 tensor). Since polarizability derivatives encode anisotropic electronic responses, Raman spectrum provides a more complex and sensitive mapping of the molecular manifold, enabling the model to resolve subtle geometric variations between conformers that the lower-rank IR descriptors may fail to distinguish.

\FloatBarrier

\section{Key contributions determining Vib2Mol performance}
The design of Vib2Conf is motivated by the fundamental disparity in information density between molecular and spectral representations. Specifically, a molecular conformation defined by atomic types and spatial coordinates is full-rank and inherently irreducible. This implies that no single atom can be removed without fundamentally altering the structural identity of the system (Figure \ref{fig:s2}). In contrast, pixel-based spectral representations are low-rank and highly redundant. As demonstrated by downsampling algorithms in signal processing, a vast majority of spectral pixels can be discarded while the underlying chemical information remains nearly intact (Figure \ref{fig:s3}).

To resolve this imbalance between high dimensionality and low information density, we implement the attentional resampler as an explicit Information Bottleneck (IB)\cite{ref23}. The optimization objective is formulated as follows.
\begin{equation}
    \min_{\theta} \left[ \mathcal{I}(X;Z) - \beta \mathcal{I}(Z;\mathcal{C}) \right]
    \label{eq:objective}
\end{equation}

In this formulation Z represents the resampled tokens. First, to reduce $\mathcal{I}\left(X;Z\right)$ and effectively filters out features irrelevant to the conformation retrieval, the number of resampled tokens is constrained to be smaller than the original spectral patches. Second, to improve $\mathcal{I}\left(Z;\mathcal{C}\right)$ and guarantee that the distilled features remain highly representative of the underlying molecular structure, resampled tokens are enforced to align with conformation features.

\begin{figure}[htbp]
    \centering
    \includegraphics[width=0.73\textwidth]{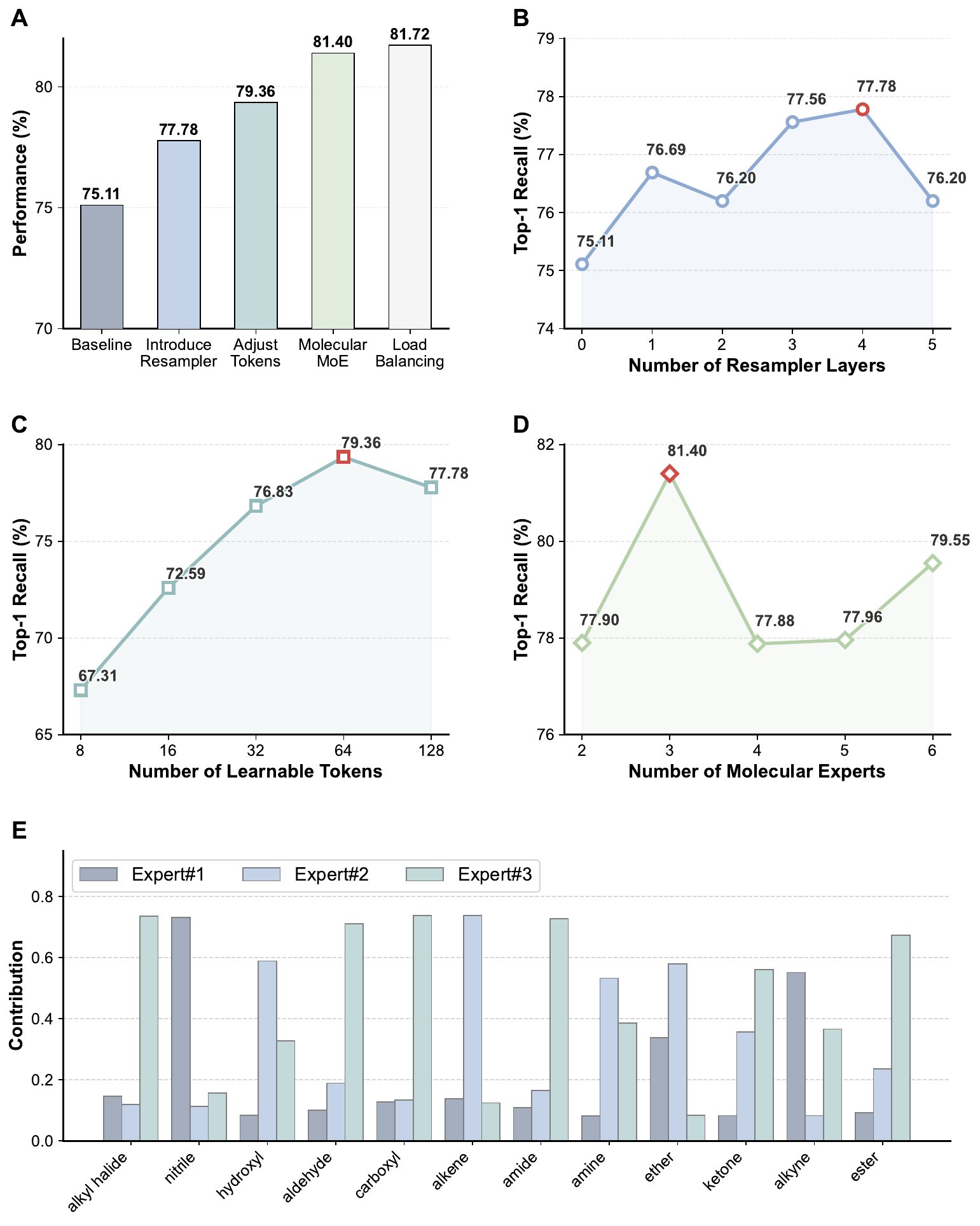}
    \caption{Ablation studies and architectural optimization of Vib2Conf. (A) Performance comparison between original and modified models. (B) Impact of the number of resampler layers. (C) Influence of the number of resampled tokens on top-1 recall. (D) Optimization of the number of experts in molecular MoE. (E) Contribution distribution across three experts.}
    \label{fig:5}
\end{figure}

Since the primary encoder is responsible for extracting features from the raw spectrum while the resampler focuses on refining those extracted features, we conducted ablation studies to determine the most effective layer for introducing the resampler and the impact of the number of resampled tokens on overall performance. These experiments identified an optimal configuration with a 2:4 ratio between the primary encoder and the attentional resampler as shown in Figure \ref{fig:5}B. Notably, the introduction of even a single resampler layer enables the model to outperform baselines that rely solely on raw spectral feature extraction. This demonstrates that refined tokens possess superior expressiveness compared to the original spectral tokens. In other words, raw spectral features contain redundant information which the resampler effectively eliminates to achieve a more compact and meaningful representation. Building upon the optimal layer configuration, we evaluated the influence of the number of resampled tokens, as shown in Figure \ref{fig:5}C. The highest performance is achieved with 64 tokens, precisely half of the original 128 spectral tokens. Reducing this count further to 32, 16, or 8 leads to a gradual performance decline, indicating that 32 tokens are insufficient to fully represent spectral information, while 64 tokens represent an ideal trade-off between representational capacity and redundancy reduction.

In contrast to the spectral distillation strategy, which necessitates the condensation of redundant information, the molecular representation requires an expanded capacity to capture the rich geometric features. Drawing inspiration from UMA\cite{ref24} and EST\cite{ref25}, we integrated the MoE into the Equiformer backbone. From a chemical perspective, the integration of MoE aims to partition the global conformational space into localized clusters. By adopting a "divide-and-conquer" strategy, the model assigns specialized experts to distinct regions of the space, thereby enhancing the precision of the spectrum-conformation retrieval.

Our ablation experiments indicate that utilizing three experts per layer yields the best results (Figure \ref{fig:5}D). As the number of experts increases beyond this point, the excessive parameterization leads to model overfitting. Conversely, an insufficient number of experts lacks the requisite capacity to capture the intricate nuances of molecular geometries. Furthermore, Figure \ref{fig:5}E illustrates the activation distribution of different experts across various functional groups, revealing that the three experts perform distinct specialized roles. Each expert exhibits a dominant preference for specific functional moieties, effectively partitioning the chemical space. In contrast, the distribution without load-balancing (Figure \ref{fig:s6}) demonstrates a significant resource collapse. In this unconstrained state, only Expert\#2 and Expert\#3 are actively engaged, while Expert\#1 remains largely dormant across most conformational inputs. This comparison underscores the necessity of the load-balancing loss in preventing expert saturation.

\FloatBarrier

\section{Discussion}
To achieve precise spectrum-conformation retrieval, we propose Vib2Conf, a deep learning model for elucidating fine-grained molecular conformation directly from vibrational spectrum. Inspired by the inherent disparity in information density between spectral and 3D molecular geometries, we introduce the attentional resampler and MoE module. The former acts as a latent bottleneck to effectively squeeze out spectral redundancies, while the latter expands the representational capacity to capture intricate conformational features. Beyond achieving state-of-the-art performance on traditional benchmarks, Vib2Conf demonstrates the exceptional capability to distinguish conformational isomers with an RMSD of approximately 1 Å, yielding a top-1 recall of 81.72\%. These results establish a robust foundation for high-throughput structural characterization and the automated identification of complex molecular ensembles in chemical and biological systems.

While Vib2Conf currently benefits from high precision from quantum chemical data, this strategy provides a robust foundation for expansion into complex experimental spectra. Once large-scale experimental datasets are available, the model will undergo rigorous empirical validation, ultimately facilitating its application in conformation-sensitive fields such as drug discovery and surface catalysis to resolve complex structural heterogeneity. Particularly, by integrating SERS-active substrate interactions with fine-tuning strategies, Vib2Conf can bridge the gap between theoretical gas-phase simulations and real-world chemical environments. This approach would enable the robust identification of adsorbed geometries on metallic surfaces, effectively extending the applicability of Vib2Conf from computational benchmarks to practical analytical chemistry, which is in progress in our group.

\begin{acknowledgement}
This work was supported by the National Natural Science Foundation (Grant No: 22227802, 22474117, 22272139 and 22595412) of China and the Fundamental Research Funds for the Central Universities (20720250080). We gratefully acknowledge the computing resources provided by the Shanghai Innovation Institute.

\end{acknowledgement}

\appendix
\section{Supplementary Information}
\setcounter{figure}{0}
\setcounter{equation}{0}
\setcounter{table}{0}

\renewcommand{\thefigure}{S\arabic{figure}}
\renewcommand{\theequation}{S\arabic{equation}}
\renewcommand{\thetable}{S\arabic{table}}

\subsection{Ablation study about load-balancing loss}
\begin{figure}[htbp]
    \centering
    \includegraphics[width=0.73\textwidth]{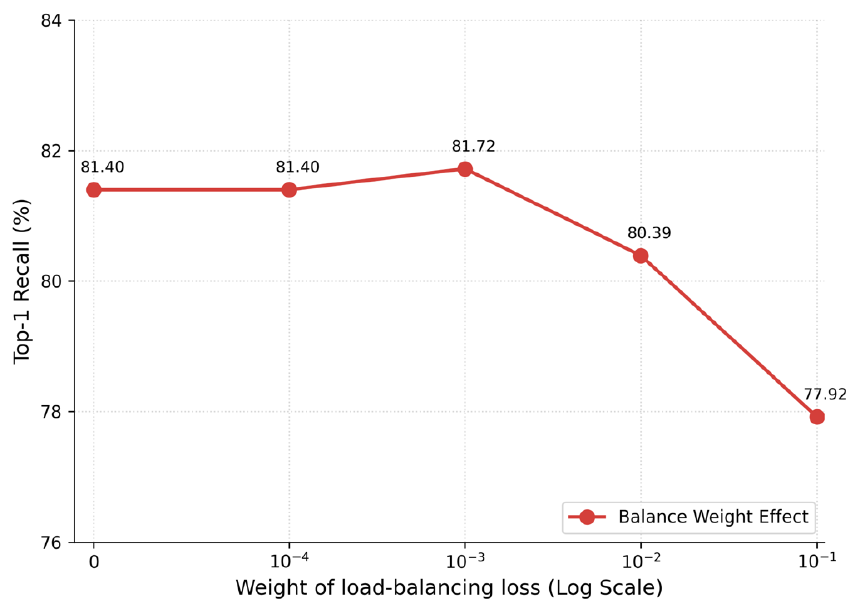}
    \caption{Effect of the load balancing weight on spectrum-conformation retrieval.}
    \label{fig:s1}
\end{figure}

While load-balancing loss is utilized to ensure a uniform workload distribution across the MoE, its intrinsic optimization objective often exhibits a divergence from the primary goal of spectrum-conformation retrieval. Excessive reliance on this auxiliary loss can inadvertently bias the learning trajectory of the model, steering it away from capturing essential geometric correlations. To mitigate this, we introduced a weighting coefficient to modulate the influence of the load-balancing term, as illustrated in Figure \ref{fig:s1}.

Our sensitivity analysis reveals that when the weight is assigned between 0.1 and 0.01, the load-balancing objective influences the optimization process, thereby degrading the performance of the retrieval task. Conversely, we identified that a weight of 0.001 yields the optimal trade-off, facilitating effective expert utilization without compromising representational accuracy. Further reducing the weight below this threshold leads to a collapse in workload balance.

\FloatBarrier

\subsection{Full-Rank Representation of Molecular Conformations}
\begin{figure}[htbp]
    \centering
    \includegraphics[width=1\textwidth]{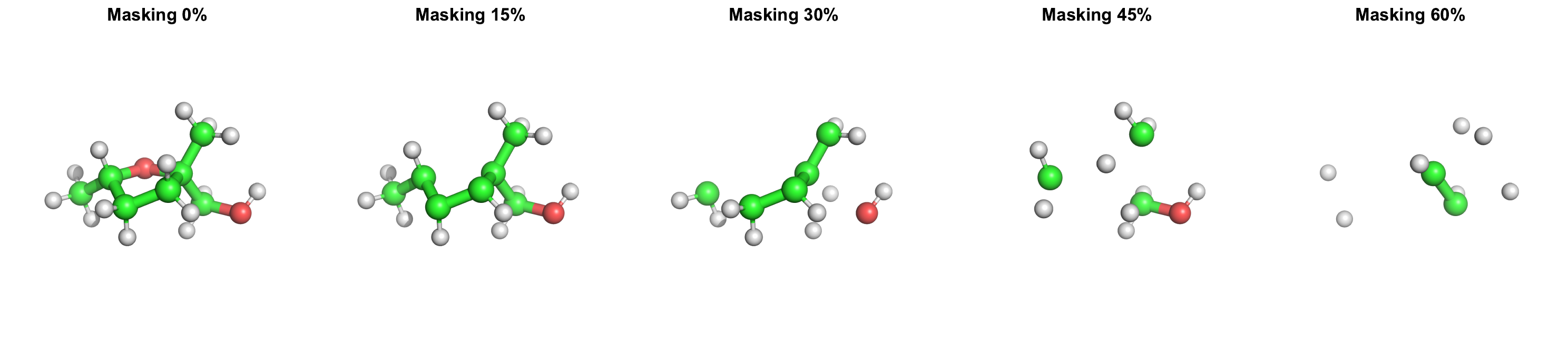}
    \caption{Molecular conformation under incremental atomic masking.}
    \label{fig:s2}
\end{figure}

A molecular conformation is defined by the joint distribution of its chemical identity and spatial geometry. Formally, we represent a conformation as a feature matrix $\mathbf{X}\in\mathbb{R}^{N\times(d+3)}$, where each row $\mathbf{x}_\mathbf{i}=[\mathbf{z}_\mathbf{i}\parallel\mathbf{r}_\mathbf{i}]$ concatenates the one-hot encoded atom type $\mathbf{z}_\mathbf{i}\in\{0,1\}^d$ with its Cartesian coordinates$\mathbf{r}_\mathbf{i}\in\mathbb{R}^{3}$.Unlike the sampling points in a 1D signal, each row in $\mathbf{X}$ is physically indispensable. The atomic type $\mathbf{z}_\mathbf{i}$ defines the nuclear charge and available valence electrons, while the coordinates $\mathbf{r}_\mathbf{i}$ determine the chemical bonding and long-range electrostatic environment. Because even the substitution of a single atom type or a sub-angstrom shift in coordinates can lead to a fundamentally different chemical species or energy minimum, the matrix $\mathbf{X}$ is inherently full-rank and non-redundant. Specifically, Figure \ref{fig:s2} illustrates the impact of randomly masking 15\%, 30\%, 45\%, and 60\% of the atoms. These results demonstrate that even minor structural omissions lead to the catastrophic disintegration of the molecular identity and the corruption of the underlying potential energy surface.

\FloatBarrier

\subsection{Spectral Redundancy and Low-Rank Projection}
In contrast, a vibrational spectrum $\mathbf{s}\in\mathbb{R}^{L}$ (where $L$ is the number of spectral bins, typically $L\ \approx1000$) exhibits profound mathematical redundancy and low-rank characteristics.
According to molecular vibration theory, any observed intensity $I\left(\nu\right)$ at wavenumber $\nu$ is a linear superposition of contributions from 3N-6 (or 3N-5) independent normal modes. Let $\phi\left(\nu-\nu_k,\gamma_k\right)$ be the line-shape function (e.g., Lorentzian) centered at frequency $\nu_k$ with broadening $\gamma_k$. The entire spectral vector $\mathbf{s}$ can be expressed as:
$$\mathbf{s}=\sum_{k=1}^{3N-6}{A_k\mathbf{b}_\mathbf{k}},\quad\mathrm{where\ }\mathbf{b}_\mathbf{k}=\left[\phi\left(\nu_1-\nu_k\right),\ldots,\phi\left(\nu_L-\nu_k\right)\right]^T$$
Here, $\mathbf{b}_\mathbf{k}\in\mathbb{R}^{L}$ serves as a basis vector for the k-th normal mode. This formulation implies that the spectral vector $\mathbf{s}$ is strictly constrained within a subspace $\mathcal{V}\subset\mathbb{R}^{L}$ spanned by the basis $\{\mathbf{b}_\mathbf{k}\}$. As a result, the maximum rank of the spectral representation is bounded by the number of vibrational degrees of freedom:
$$\mathrm{rank}\left(\mathbf{s}\right)\le\min{\left(L,3N-6\right)}$$
Given $L\ \gg3N-6$ for a medium-sized molecule, the spectral matrix is intrinsically low-rank and singular.

\begin{figure}[htbp]
    \centering
    \includegraphics[width=1\textwidth]{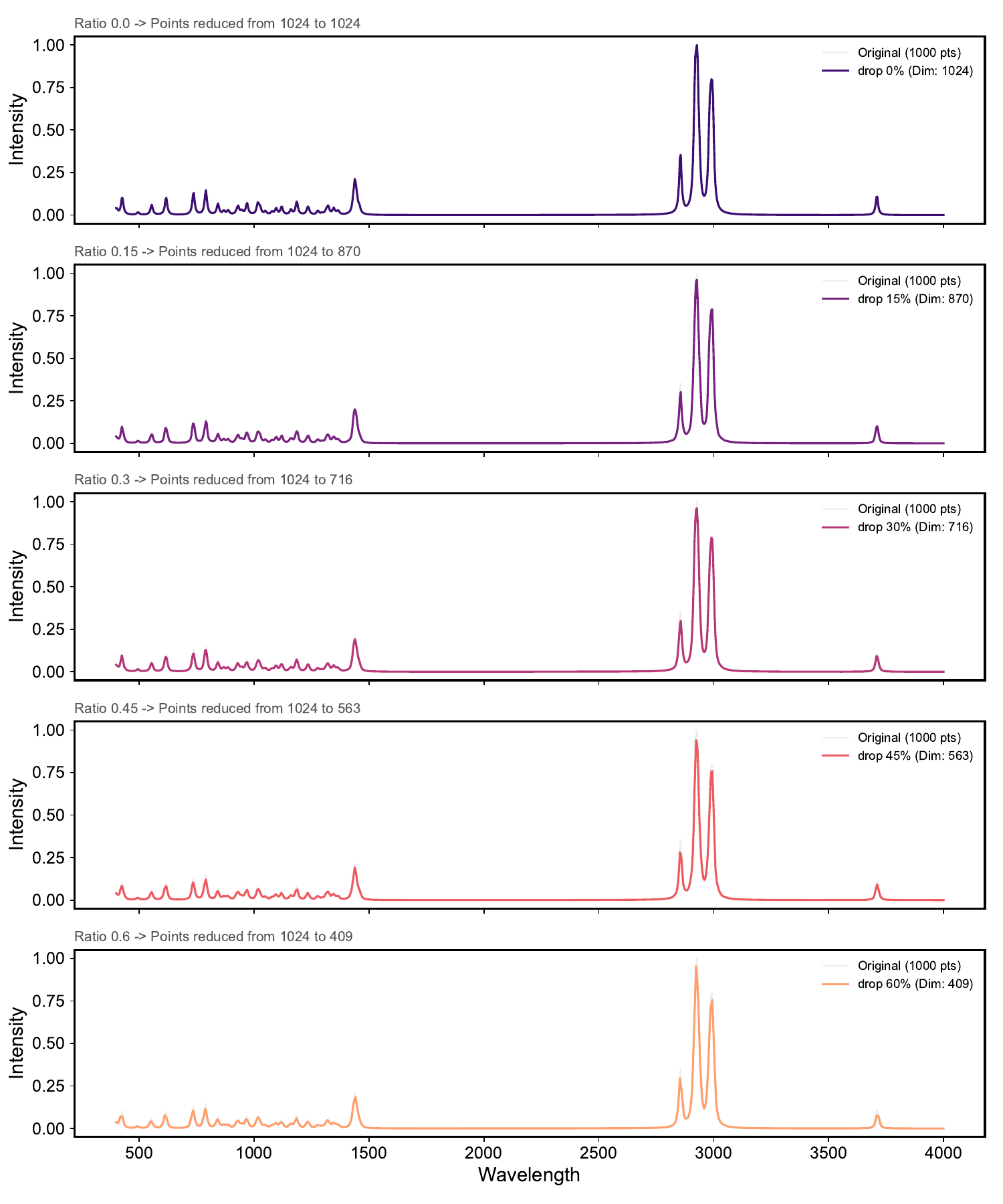}
    \caption{Spectra under incremental downsampling. The method employed here for downsampling is average pooling.}
    \label{fig:s3}
\end{figure}
\FloatBarrier

\subsection{Feature Distillation via Cross-Attention}
The transition from the initial spectral representations to the distilled latent tokens is mediated by cross-attention, which acts as a trainable information bottleneck. Let $H_p \in \mathbb{R}^{128\times D}$ denote the primary spectral features extracted by the initial encoder, where 128 represents the number of input patches. To distill these redundant signals, we initialize a set of learnable query tokens $T_{init}\in\mathbb{R}^{64\times D}$. The cross-attention operation $\mathrm{Cross-Attention}\left(\cdot\right)$ transforms the primary features into a refined latent space:
$$H_r=\mathrm{Softmax}\left(\frac{Q_rK_p^T}{\sqrt{d_k}}\right)V_p$$
In this formulation, the queries $Q_r=T_{init}W^Q$ are derived from the learnable tokens, representing the model's active "probes" for essential vibrational signatures. The keys $K_p=H_pW^K$ and values $V_p=H_pW^V$ are projected from the primary spectral features.
By using a reduced set of 64 queries to aggregate information from 128 primary patches, the mechanism effectively filters out the redundancy inherent in $H_p$, thereby ensuring efficient alignment during contrastive learning.

\subsection{Equations}
\begin{equation}
    L_{contrastive}=\frac{1}{2}\left(L_{spectrum}+L_{structure}\right)=-\frac{1}{2B}\left(\sum_{i=1}^{B}\log{\left(p_i\right)}+\sum_{j=1}^{B}\log{\left(q_j\right)}\right)
    \label{eqn:s1}
\end{equation}

Expanding this with the similarity scores, the formulation becomes:
\begin{equation}
    L_{contrastive}=-\frac{1}{2B}\left(\sum_{i=1}^{B}\log{\frac{\exp{\left(s_{i,i}/\tau\right)}}{\sum_{j=1}^{B}\exp{\left(s_{i,j}/\tau\right)}}}+\sum_{j=1}^{B}\log{\frac{\exp{\left(s_{j,j}/\tau\right)}}{\sum_{i=1}^{B}\exp{\left(s_{i,j}/\tau\right)}}}\right)
    \label{eqn:s2}
\end{equation}

Where $B$ denotes the size of mini-batch, $p_i$ and $q_j$ represent the predicted probabilities of correct alignment of spectrum-to-structure, and structure-to-spectrum. These probabilities are calculated based on the similarity scores $s_{i,j}$ and scaled by a temperature parameter $\tau$.

\subsection{Tables}

\begin{table}[htbp]
  \centering
  \caption{Hyperparameters for training.}
    \begin{tabular}{lc}
    \toprule
    Training epochs & 150 \\
    Learning rate & 0.00005 \\
    Batch size & $128\times4 $\\
    Number of spectral encoder layers & 6 \\
    Number of molecular encoder layers & 6 \\
    Number of spectral encoder channels & 512 \\
    Number of molecular encoder channels & $128\times0e+64\times1e+32\times2e$ \\
    Number of spectral encoder heads & 8 \\
    Number of molecular encoder heads & 4 \\
    Cut-off radius (Å) & 5 \\
    \bottomrule
    \end{tabular}%
  \label{tab:hyperparams}%
\end{table}%

\begin{table}[htbp]
  \centering
  \caption{Spectrum-conformation retrieval performance on the QM9S test set, reported as Top-1 Recall (\%).}
    \begin{tabular}{cccc}
    \toprule
     & Raman & IR & IR+Raman \\
    \midrule
    Vib2Mol & 97.18 & 95 & 98.42 \\
    VibraCLIP & 97.07 & 95.23 & 98.21 \\
    SMEN & 97.89 & 97.04 & - \\
    Vib2Conf & \textbf{97.37} & \textbf{97.57} & \textbf{98.50} \\
    \bottomrule
    \end{tabular}%
  \label{tab:qm9s}%
\end{table}%

\begin{table}[htbp]
  \centering
  \caption{Spectrum-conformation retrieval performance on the VB-Mols test set, reported as Top-1 Recall (\%).}
    \begin{tabular}{cccc}
    \toprule
     & Raman & IR & IR+Raman \\
    \midrule
    Vib2Mol & 93.2 & 91.38 & 94.26 \\
    VibraCLIP & 91.21 & 89.01 & 95.91 \\
    SMEN & 95.43 & 94.02 & - \\
    Vib2Conf & \textbf{96.95} & \textbf{97.23} & \textbf{98.55} \\
    \bottomrule
    \end{tabular}%
  \label{tab:vbmols}%
\end{table}%

\begin{table}[htbp]
  \centering
  \caption{Spectrum-conformation retrieval performance on the QMe14S test set, reported as Top-1 Recall (\%).}
    \begin{tabular}{cccc}
    \toprule
     & Raman & IR & IR+Raman \\
    \midrule
    Vib2Mol & 95.35 & 94.45 & 95.96 \\
    VibraCLIP & 88.78 & 87.16 & 96.00 \\
    SMEN & 93.96 & 93.81 & - \\
    Vib2Conf & \textbf{95.58} & \textbf{96.73} & \textbf{97.90} \\
    \bottomrule
    \end{tabular}%
  \label{tab:qme14s}%
\end{table}%

\begin{table}[htbp]
  \centering
  \caption{Multi-modal spectrum-conformation retrieval performance on three test sets, reported as Top-1 Recall (\%).}
    \begin{tabular}{cccc}
    \toprule
     & Raman & IR & IR+Raman \\
    \midrule
    QM9S & 97.37 & 97.57 & \textbf{98.50} \\
    VB-Mols & 96.95 & 97.23 & \textbf{98.55} \\
    QMe14S & 95.58 & 96.73 & \textbf{97.90} \\
    \bottomrule
    \end{tabular}%
  \label{tab:multi-modal}%
\end{table}%

\begin{table}[htbp]
  \centering
  \caption{Spectrum-conformation retrieval performance on the VB-Confs test set, reported as Top-1 Recall (\%).}
    \begin{tabular}{cccc}
    \toprule
     & Perfectly correct & Partially correct & Incorrect \\
    \midrule
    Raman & 81.72 & 18.22 & 0.06 \\
    IR & 74.52 & 25.01 & 0.47 \\
    Raman+IR & 82.06 & 17.92 & 0.02 \\
    \bottomrule
    \end{tabular}%
  \label{tab:confs}%
\end{table}%

\FloatBarrier

\subsection{Figures}

\begin{figure}[htbp]
    \centering
    \includegraphics[width=0.73\textwidth]{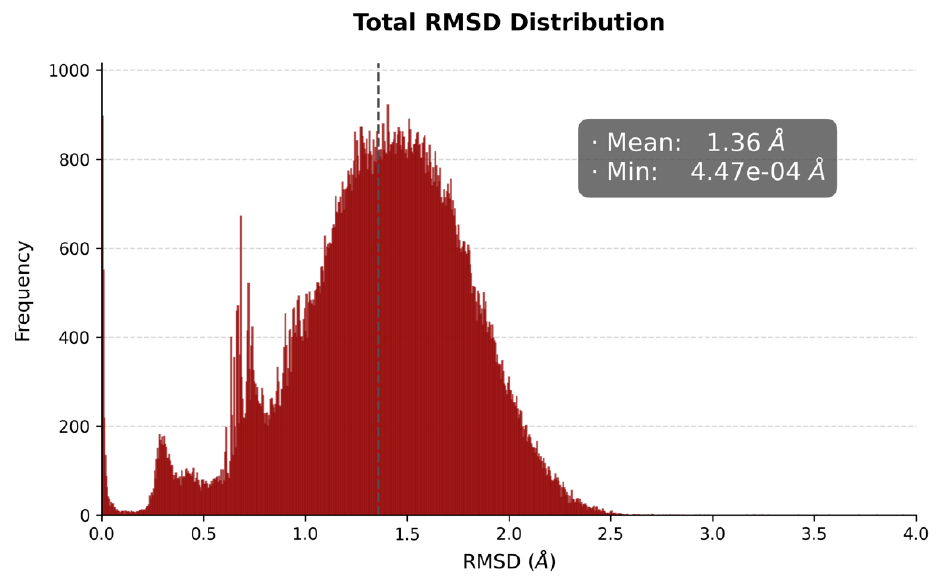}
    \caption{RMSD distribution of VB-Confs.}
    \label{fig:s4}
\end{figure}

\begin{figure}[htbp]
    \centering
    \includegraphics[width=0.73\textwidth]{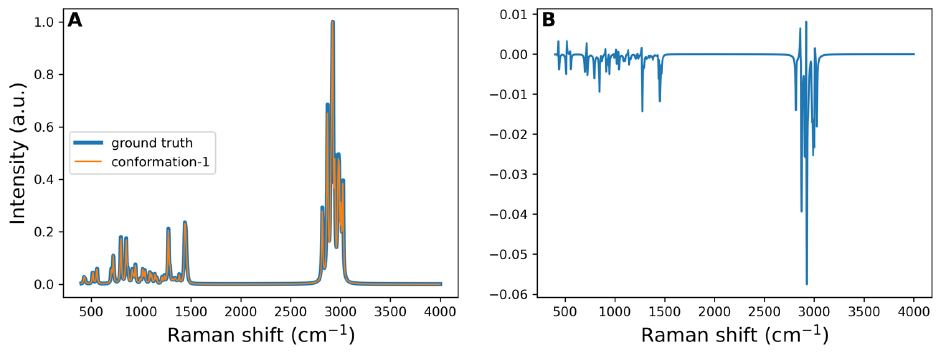}
    \caption{(A) Original spectra and (B) difference spectrum of ground truth and conformation\#1.}
    \label{fig:s5}
\end{figure}

\begin{figure}[htbp]
    \centering
    \includegraphics[width=0.73\textwidth]{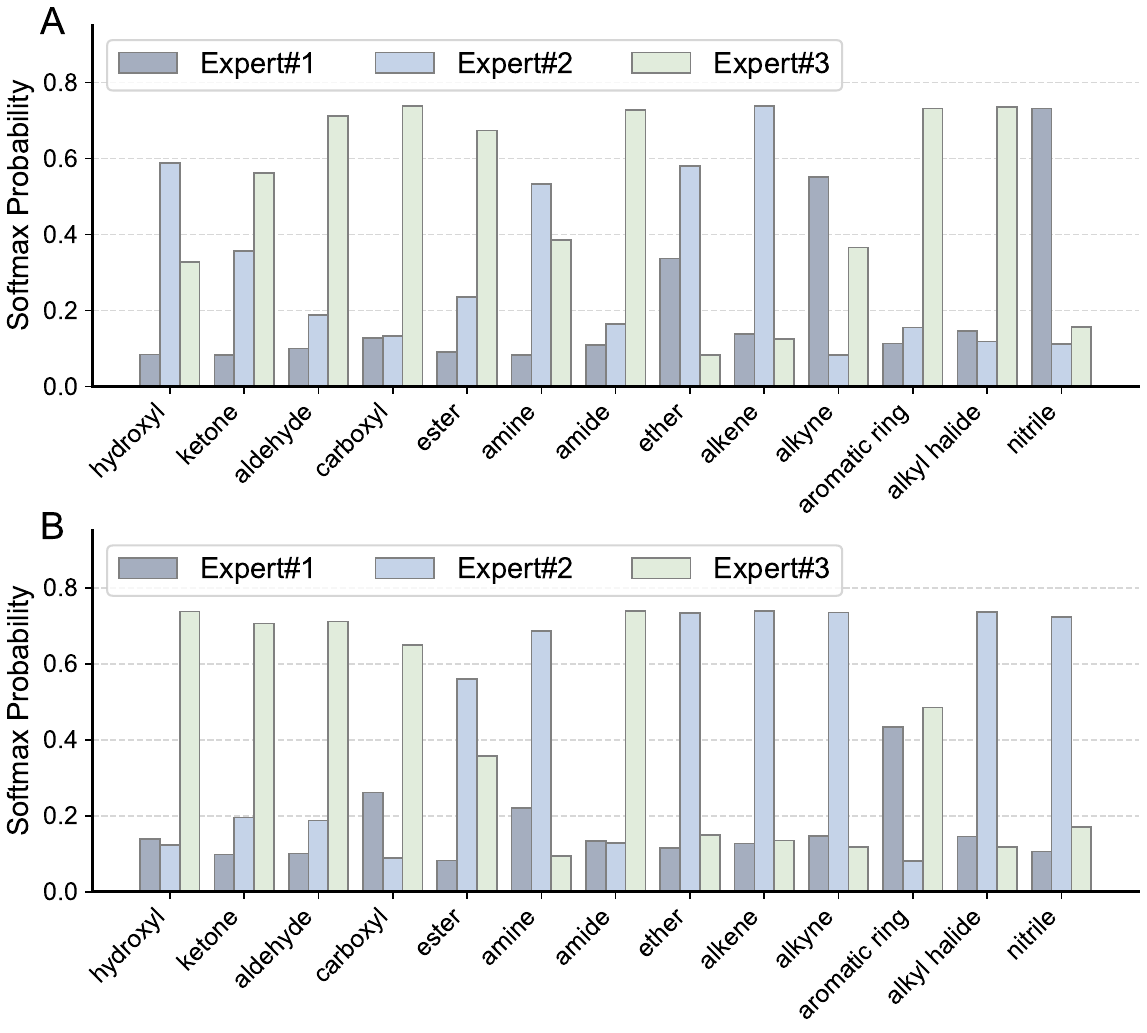}
    \caption{Functional group-specific activation distribution of individual experts (A) with and (B) without load-balancing loss.}
    \label{fig:s6}
\end{figure}
\FloatBarrier

\bibliography{achemso-demo}

\end{document}